\documentclass[11pt,twoside]{article}
\usepackage{amsmath}
\usepackage{amssymb}
\usepackage{amscd}
\usepackage{graphicx}
\usepackage{epstopdf}
\usepackage[usenames]{color}
\def\EndBox#1{
	\hskip0.1em\hfill\null\ \null\nobreak\hfill\kern3pt
		\hbox{$\scriptstyle #1$} \smallbreak}
\def\qed{\EndBox{\square}}
\newtheorem{proposition}{Proposition}[section]
\newtheorem{theorem}{Theorem}[section]
\DeclareFontFamily{U}{UWCyr}{}
\DeclareFontShape{U}{UWCyr}{m}{n}{%
  <5> <6> <7> <8> <9>
  <10> <10.95> <12> <14.4> <17.28> <20.74> <24.88> wncyr10
  }{}
\DeclareFontShape{U}{UWCyr}{m}{it}{%
  <5> <6> <7> <8> <9>
  <10> <10.95> <12> <14.4> <17.28> <20.74> <24.88> wncyi10
  }{}
\DeclareFontShape{U}{UWCyr}{m}{sc}{%
  <5> <6> <7> <8> <9>
  <10> <10.95> <12> <14.4> <17.28> <20.74> <24.88> wncysc10
  }{}
\DeclareFontShape{U}{UWCyr}{b}{n}{%
  <5> <6> <7> <8> <9>
  <10> <10.95> <12> <14.4> <17.28> <20.74> <24.88> wncyb10
  }{}
\DeclareMathAlphabet{\cyrm}{U}{UWCyr}{m}{n}
\newsavebox{\remendsymbol}
\savebox{\remendsymbol}[6pt]{
{\begin{picture}
	(10,10)\linethickness{2pt}\put(0,0){\line(1,0){8}}
			\put(7,0){\line(0,1){8}} \end{picture}}  }
\newcommand{\remend}{{\usebox{\remendsymbol}}}
\newcommand{\rembox}{\EndBox{\boldsymbol{\remend}}}

\newcommand{\proof}{{\sc proof:~}}
\newcommand{\remark}{\smallbreak\noindent{\bf Remark.}}
\renewcommand{\a}{\alpha}

\newcommand{\g}{\gamma}
\renewcommand{\d}{\delta}
\newcommand{\e}{\varepsilon}
\newcommand{\be}{{\bar\varepsilon}}
\newcommand{\ze}{{\zeta}}
\newcommand{\Th}{\Theta}
\renewcommand{\l}{\lambda}
\newcommand{\m}{\mu}
\newcommand{\n}{\nu}
\renewcommand{\r}{\rho}
\newcommand{\s}{\sigma}
\newcommand{\Om}{{\Omega}}
\newcommand{\h}{\hbar}
\newcommand{\de}{\partial}
\newcommand{\oh}{\tfrac{1}{2}}
\newcommand{\ih}{\tfrac{\iO}{2}}
\newcommand{\cj}[1]{\overline{#1}}
\newcommand{\lin}{{\scriptscriptstyle\bigstar}}

\renewcommand{\.}{{\scriptstyle\boldsymbol{\dot{}}}}
\newcommand{\td}{\tilde}
\newcommand{\br}{\breve}
\newcommand{\up}{{\scriptscriptstyle\uparrow}}

\newcommand{\sbot}{{\scriptscriptstyle\bot}}
\newcommand{\spar}{{\scriptscriptstyle\|}}
\newcommand{\ost}[1]{\overset{{}_{{\,}_*}}#1}
\newcommand{\fl}{\flat}
\newcommand{\B}{{\boldsymbol{B}}}
\newcommand{\E}{{\boldsymbol{E}}}
\newcommand{\F}{{\boldsymbol{F}}}
\renewcommand{\H}{{\boldsymbol{H}}}
\newcommand{\Q}{{\boldsymbol{Q}}}

\newcommand{\T}{{\boldsymbol{T}}}
\newcommand{\U}{{\boldsymbol{U}}}
\newcommand{\Uc}{\cj{\U}}
\newcommand{\Ua}{\cj{\U}{}^\lin}
\newcommand{\Ul}{\U{}^\lin}
\newcommand{\W}{{\boldsymbol{W}}}

\newcommand{\Lie}{\boldsymbol{\mathfrak{L}}}
\newcommand{\Ug}{\mathrm{U}}
\newcommand{\SO}{\mathrm{SO}}
\newcommand{\SU}{\mathrm{SU}}
\newcommand{\CC}{{\mathbb{C}}}
\newcommand{\II}{{\mathbb{I}}}
\newcommand{\LL}{{\mathbb{L}}}
\newcommand{\MM}{{\mathbb{M}}}

\newcommand{\RR}{{\mathbb{R}}}
\newcommand{\TT}{{\mathbb{T}}}
\newcommand{\UU}{{\mathbb{U}}}
\newcommand{\Hcal}{{\mathcal{H}}}
\newcommand{\Pcal}{{\mathcal{P}}}
\newcommand{\Ycal}{{\mathcal{Y}}}

\newcommand{\DC}{{\boldsymbol{\mathcal{D}}}}
\newcommand{\HC}{{\boldsymbol{\mathcal{H}}}}
\newcommand{\JC}{{\boldsymbol{\mathcal{J}}}}
\newcommand{\LC}{{\boldsymbol{\mathcal{L}}}}
\newcommand{\DCo}{\DC_{\!\circ}}
\newcommand{\End}{\operatorname{End}}
\newcommand{\Tr}{\operatorname{Tr}}
\newcommand{\Id}[1]{{1\!\!1}\!{}_{#1}{}}
\newcommand{\id}{{1\!\!1}}
\newcommand{\CO}{\mathrm{C}}
\newcommand{\dO}{\mathrm{d}}

\newcommand{\jO}{\mathrm{j}}
\newcommand{\JO}{\mathrm{J}}

\newcommand{\TO}{\mathrm{T}}
\newcommand{\TS}{\TO^{*}\!}
\newcommand{\VO}{\mathrm{V}}
\newcommand{\VS}{\VO^{*}\!}
\newcommand{\dx}{\dO\xx}

\newcommand{\iO}{\mathrm{i}}
\newcommand{\pr}[1]{\operatorname{pr}_{#1}}
\newcommand{\na}{\nabla\!}
\newcommand{\ten}[1]{\operatorname*{\otimes}_{\!{\scriptscriptstyle #1}} }
\newcommand{\cart}[1]{\operatorname*{\times}_{\!{\scriptscriptstyle #1}} }
\newcommand{\dir}[1]{\operatorname*{\oplus}_{\!{\scriptscriptstyle #1}} }
\newcommand{\we}{{\,\wedge\,}}
\newcommand{\weu}[1]{{\wedge^{\!#1}}}
\newcommand{\pint}{\,\mathord{\lrcorner}\,}
\newcommand{\comp}{\mathbin{\raisebox{1pt}{$\scriptstyle\circ$}}}
\newcommand{\tn}{{\,\otimes\,}}
\newcommand{\bang}[1]{{\langle#1\rangle}}
\newcommand{\Ii}[2]{{}^{#1}_{\phantom{#1}\!#2}}
\newcommand{\iI}[2]{{}_{#1}^{\phantom{#1}\!#2}}
\newcommand{\iIi}[3]{{}_{#1\phantom{#2}\!\!#3}^{\phantom{#1}\!#2}}
\newcommand{\sA}{{\scriptscriptstyle A}}
\newcommand{\sB}{{\scriptscriptstyle B}}
\newcommand{\sC}{{\scriptscriptstyle C}}
\newcommand{\cA}{{\sA\.}}
\newcommand{\cB}{{\sB\.}}

\newcommand{\zeA}{{\zeta_\sA}}
\newcommand{\dezA}{\de\zz_\sA}

\newcommand{\zzA}{\zz^\sA}
\newcommand{\zzB}{\zz^\sB}
\newcommand{\ee}{{\mathsf{e}}}
\newcommand{\xx}{{\mathsf{x}}}

\newcommand{\zz}{{\mathsf{z}}}
\newcommand{\ie}{i.e$.$}
\newcommand{\eg}{e.g$.$}
\newcommand{\hm}{\phantom{-}}
\newcommand{\sst}{\scriptscriptstyle}
\newcommand{\into}{\hookrightarrow}
\renewcommand{\le}{\left}
\newcommand{\ri}{\right}

\newcommand{\ii}{{\scriptstyle{\II}}}
\newcommand{\Ch}{\cyrm{Ch}}
\newcommand{\JE}{{\JO\E}}
\newcommand{\TE}{{\TO\E}}
\newcommand{\VE}{{\VO_{\!}\E}}
\newcommand{\JU}{{\JO\U}}
\newcommand{\TU}{{\TO\U}}
\newcommand{\VU}{{\VO\U}}

\newcommand{\bbra}[1]{[\![#1]\!]}
\newcommand{\chY}{\check Y}
\newcommand{\dAB}{\d\Ii\sA\sB}
\newcommand{\Pau}{{\scriptstyle\Sigma}}
\newcommand{\tm}{{\mathfrak{t}}}
\newcommand{\nat}{\natural}
\newcommand{\hb}{\hbox}
\newcommand{\Aie}{\boldsymbol{\mathfrak{A}}}
\newcommand{\tri}{\triangleright}
\newcommand{\diveta}{\mathop{\mathrm{div}_\eta}}
\title{Hermitian vector fields and covariant quantum mechanics\\
of a spin particle}
\date{{\it{\small 9 September 2009}} }
\author{Daniel Canarutto
\\[6pt]
{\small\it Dipartimento di Matematica Applicata ``G. Sansone'', }\\
{\small\it Via S. Marta 3, 50139 Firenze, Italia}}
\begin{document}
\bibliographystyle{alpha}
\maketitle
\begin{abstract}\noindent
In the context of Covariant Quantum Mechanics for a spin particle,
we classify the ``quantum vector fields'',
\ie\ the projectable Hermitian vector fields of a complex bundle
of complex dimension 2 over spacetime.
Indeed, we prove that the Lie algebra of quantum vector fields
is naturally isomorphic to a certain Lie algebra of functions
of the classical phase space, called ``special phase functions''.
This result provides a covariant procedure to achieve the quantum operators
generated by the quantum vector fields
and the corresponding observables described by the special phase functions.
\end{abstract}

\noindent
2000 MSC:
17B66, %Lie algebras of vector fields and related (super) algebras
17B81, %Applications to physics
53C27, %Spin and Spin$^c$ geometry
53B35, %Hermitian and KŠhlerian structures
53C50, %Lorentz manifolds, manifolds with indefinite metrics
81S10, %Geometry and quantization, symplectic methods
81R25. %Spinor and twistor methods

\noindent
Keywords:
Hermitian vector fields,
special functions,
covariant quantum mechanics,
spin particle,
Galileian spacetime.

\bigbreak\noindent
{\sc Acnowledgments}\\
The author is grateful
to Marco Modugno
(Dipartimento di Matematica Applicata ``G. Sansone'', Firenze)
and Josef Jany\v{s}ka
(Department of Mathematics, Masaryk University, Czech Republic)
for many remarks and discussions,
and important clarifications about the scalar-particle theory.

\tableofcontents

%------------------------------------------------------------------------------%
\section*{Introduction}
%------------------------------------------------------------------------------%

We deal with a geometric formulation of quantum mechanics
on a curved spacetime with absolute time
(``curved Galileian spacetime'')
which was introduced by M.Modugno and A.~Jadczyk~\cite{JadMod92,JadMod94}
and further developed by several authors~\cite{CJM95,JadJanMod98,
Jan95a,JanMod97,JanMod02b,JanMod02c,JM06,JanMod05p1,
JanModSal02,ModSalTol05,ModTejVit00,
SalVit00,Vit96,Vit99}.
This approach, named \emph{Covariant Quantum Mechanics} (CQM),
uses differential geometric notions such as
jets of fibered manifolds, linear and non-linear connections,
cosymplectic forms and Fr\"olicher smooth spaces;
it has analogies with
Geometric Quantization~\cite{AbrMar78,Gar79,Got98,Kos70,Sni80,Sou70,Woo92}
and with other formulations such as the approach due to Kuchar~\cite{Kuc80}
and the approach by C.~Duval, K\"unzle et al.~\cite{DuvBurKunPer85, DuvKun84}
in the Bargmann framework.

On the other hand, CQM has several distinctive features;
it produces an effective procedure for the introduction of quantum operators,
and overcomes certain difficulties typical of Geometric Quantization---%
such as the problems related to polarizations
and to the quantum energy operator;
moreover it reproduces, in the flat case, the whole standard quantum mechanics.

Within CQM one finds a precise criterium yielding quantum operators;
these are in one-to-one natural correspondence
with certain ``special functions'',
which on turn constitute a Lie algebra
different from the usual Poisson algebra.
\remark~The Poisson bracket is defined for any pair of phase-space functions,
while special function are just a ``small'' subset of these;
on the other hand, the space of special functions is not closed
with respect to the Poisson bracket,
but it is closed with respect to a special bracket,
which is defined as the Poisson bracket \emph{plus} a certain covariant term.
The special bracket coincides with the Poisson bracket in the particular case
when both factors are special functions fulfilling a further condition
(being affine functions on the fibres of the phase space).
Hence this subspace of all special functions constitutes a Lie subalgebra
of both the Poisson algebra and the special algebra.\footnote{
In a sense the special bracket is analogous to the Jacobi bracket,
though they still are different operations~\cite{JM09}.} 
Now the special bracket (and everything related to it)
is relevant exactly in the case of quadratic functions,
namely functions containing the \emph{energy}.
This is one of the basic points which distinguish between CQM
and standard Geometric Quantization or other geometric approaches:
the energy function can be treated on the same footing
as all other quantizable functions,
without \emph{ad hoc} tricks and having to deal with ordering problems.~\remend
\smallbreak

One of the relatively recent developments~\cite{JM06} of CQM
has been exactly a deeper understanding of the strict relation existing
among ``special functions'',
corresponding to a distinguished class of observables,
and Hermitian vector fields on the fundamental complex bundle of the theory
(on turn, Hermitian vector fields yield quantum operators).
This paper aims at extending those results
(stated in the case when the bundle's fibres are 1-dimensional)
to the case of a particle with spin (complex 2-dimensional fibres).
While the spin case was already studied in a previous article~\cite{CJM95},
the algebra of special observable functions
was introduced there in a rather ``by-hand'' fashion.
Now it can be shown to arise from quite natural geometric constructions,
and to need further corrections due to the interplay between
the geometry of the quantum structures and the underlying
classical spacetime geometry.

The first section of this paper is devoted to preliminary results
about Hermitian vector fields on Hermitian vector bundles,
and their relation to Hermitian connections.

The second section contains a sketch of the
Galileian spacetime geometry
and related classical particle mechanics underlying CQM.
The extension to the case of a classical particle with spin
is also considered.
The Lie algebra of special functions for the scalar case
is briefly  reviewed,
and its extension to the spin case is presented.

In the third section we introduce the geometric framework
needed for the treatment of a quantum spin particle in the Galileian context:
spin bundle, Pauli map, spin connection.
Moreover, a comparison is made between this framework
and the framework for the Dirac equation in General Relativity.

In the fourth and last section we introduce the phase-quantum bundle
and the phase-quantum connection as a natural extension
of the scalar particle case;
we study the classification of Hermitian vector fields
and their relations with the special functions.
Finally we show how quantum operators and the Pauli equation
on a Galileian background arise.

This paper does not contain a complete treatment of CQM,
with its rich geometric structure and delicate points.
We rather sketch some of the main features
in order to frame our results.
For a more detailed understanding of the theory the reader is advised to look
at some of the above quoted papers~\cite{JanMod02b, JanMod02c, JM06}.

%==============================================================================%
\section{Hermitian vector fields}
\label{s:Hermitian vector fields}

This section contains some preliminary mathematical results.

%--------------
\subsection{Hermitian Lie algebra}\label{ss:Hermitian Lie algebra}
If $\U$ is a complex vector space
of finite dimension $n$
then we indicate by $\Ul$, $\Uc$ and $\Ua$
its dual, conjugate and anti-dual spaces, respectively.
A Hermitian metric on $\U$ is a positive-defined tensor
\hb{$h\in\Ua\tn\Ul$} which is preserved by
anri-transposition~\cite{C00b}.
The space $\Lie\subset\U\tn\Ul\equiv\End\U$
consisting of all $h$-antihermitian endomorphisms
is a Lie subalgebra of $\End\U$
(the product of endomorphisms coinciding with ordinary commutator).
Actually, $\Lie$ is the Lie algebra of Lie group $\Ug(\U,h)$,
consisting of all $h$-unitary automorphisms of $\U$.
If we fix an $h$-orthonormal basis of $\U$
then we get isomorphisms $\Ug(\U,h)\cong\Ug(n)$\,,
and between $\Lie$ and the Lie algebra of $\Ug(n)$\,.

The Lie subalgebra $\Lie_0\subset\Lie$
consisting of all antihermitian \emph{traceless} endomorphisms
is the Lie algebra of the subgroup $\SU(\U,h)\subset\Ug(\U,h)$,
consisting of all unitary automorphisms whose determinant equals $+1$
($\Lie_0$ vanishes if $n=1$).

%--------------
\subsection{Projectable vector fields}\label{ss:Projectable vector fields}
We consider a finite-dimensional complex vector bundle $\pi:\U\to\E$
over the real manifold $\E$.
We indicate by $\bigl(\xx^\l\bigr)$ a local coordinate chart on $\E$,
and by $\bigl(\zeA\bigr)$ a local frame of $\U$\,.
We use fibered coordinates
\hb{$\bigl(\xx^\l,\zz^\sA\bigr):\U\to\RR^m\times\CC^n$},
where the linear fiber coordinates $\bigl(\zz^\sA\bigr)$
constitute the dual frame of $\bigl(\zeA\bigr)$\,.
Moreover we indicate by $\bigl(\dezA\bigr)$
the induced frame\footnote{
We indicate the tangent, vertical and fist-jet prolongtion functors
as $\TO$, $\VO$ and $\JO$\,, respectively.
Taking into account the natural isomorphism \hb{$\VU\cong\U\cart{\E}\U$}
we actually have $\dezA\equiv\zeA$\,.} 
of $\VU\to\U$\,.

A vector field \hb{$Y:\U\to\TU$} is said to be \emph{projectable}
if there exists a vector field \hb{$X{:}\,E\,{\to}\,\TE$}
such that the following diagram commutes:
$$\begin{CD}
\U @>{Y}>> \TU \\
@VVV  @VVV \\
\E @>>{X}> \TE
\end{CD}$$
In particular we say that $Y$ is \emph{linear projectable}
if it is a linear morphism over $X$\,;
then its coordinate expression is
$$Y=X^\l\,\de\xx_\l+(Y\Ii\sA\sB\,\zzB\,)\,\dezA~,\quad
X^\l:\E\to\RR~,~~
Y\Ii\sA\sB:\E\to\CC~.$$

A short calculation shows that
the Lie bracket of two vector fields $Y$ and $Y'$,
respectively projectable over $X$ and $X'$,
is projectable over $[X,X']$\,;
if morever $Y$ and $Y'$ are linear then also $[Y,Y']$
turns out to be linear,
and has the coordinate expression
\begin{align*}
[Y,Y']&=(X^\l\,\de_\l X'^\m-X'^\l\,\de_\l X^\m)\,\de\xx_\m+{}
\\[6pt]
&\quad+(X^\l\,\de_\l Y'\Ii\sA\sB-X'^\l\,\de_\l Y\Ii\sA\sB
+Y'\Ii\sA\sC\,Y\Ii\sC\sB-Y\Ii\sA\sC\,Y'\Ii\sC\sB)\,\zzB\,\dezA~.
\end{align*}
Thus local linear projectable vector fields
constitute a sheaf of Lie algebras
(a subsheaf of the sheaf of all vector fields $\U\to\TU$
together with the standard Lie bracket).

%--------------
\subsection{Hermitian vector fields}
\label{ss:Hermitian vector fields}

Let $\psi:\E\to\U$ be a (local) section,
and write its coordinate expression as \hb{$\psi=\psi^\sA\,\zeA$}\,,
\hb{$\psi^\sA:\E\to\CC$}\,.
Recalling the natural isomorphism
\hb{$\U\cart{\E}\U\cong\VU\into\TU$}
we define the vertical vector field
$$\td\psi:\E\to\U\cart{\E}\U\cong\VU\into\TU:
u\mapsto\td\psi(u):=\bigl(u,\psi(\pi u)\bigr)~,$$
which has the coordinate expression $\td\psi=\psi^\sA\,\dezA$\,.
If $Y:\U\to\TU$ is linear projectable then the vertical vector field
$$[Y,\td\psi]=(X^\l\,\de_\l\psi^\sA-Y\Ii\sA\sB\,\psi^\sB)\,\dezA:
\U\to\VU\cong\U\cart{\E}\U$$
yields, through projection onto the second cartesian factor,
a section $Y.\psi$ with coordinate expression
$$Y.\psi=(X^\l\,\de_\l\psi^\sA-Y\Ii\sA\sB\,\psi^\sB)\,\zeA:\E\to\U~.$$

Now we assume that the fibers of $\U\to\E$ be smoothly endowed
with a Hermitian metric,
namely that we have a smooth section \hb{$h:\E\to\Ua\ten{\E}\Ul$}\,.
Then we say that a linear projectable vector field $Y$ is \emph{Hermitian}
if for any two sections \hb{$\phi,\psi:\E\to\U$} we have\footnote{
The Lie derivative along linear projectable vector fields
can be extended, in a natural way,
to the tensor algebra of $\U\to\E$,
including tensors with indices of all types
(covariant and contravariant, dotted and non-dotted).
Then the condition that $Y$ be Hermitian
can be also directly expressed as $Y.h=0$\,.} 
$$X.(h(\phi,\psi))=h(Y.\phi,\psi)+h(\phi,Y.\psi)~.$$
In coordinates the above condition reads
\hb{$h_{\cA\sB}\,\bar Y\Ii\cA\cB+h_{\cB\sA}\,Y\Ii\sA\sB=
X^\l\,\de_\l h_{\cB\sB}$}\,.
In the sequel we'll use $h$-orthonormal local fiber coordinates,
hence the same condition becomes
$$\d_{\cA\sB}\,\bar Y\Ii\cA\cB+\d_{\cB\sA}\,Y\Ii\sA\sB=0~;$$
namely, the matrix $\bigl(Y\Ii\sA\sB\bigr)$ is \emph{anti-Hermitian}.

In particular, if \hb{$Y:\U\to\VU\cong\U\cart{\E}\U$}
then $Y$ can also be seen as a section \hb{$\E\to\End\U$}\,;
in that case the Hermicity condition means \hb{$Y:\E\to\Lie$}\,,
namely $Y$ is a section
of the subbundle of \hb{$\End\U$} constituted by all traceless anti-Hermitian
endomorphisms of the fibers of \hb{$\U\to\E$}\,.

\begin{proposition}
Let $Y,Y':\U\to\TU$ be linear vector fields,
projectable over $X,X':\E\to\TE$.
Then their Lie bracket $Y,Y'$ is a linear vector field,
projectable over $[X,X']$\,.
Actually by a straightforward coordinate calculation one finds
\begin{align*}
[Y,Y']&=
(X^\l\,\de_\l X'^\m-X'^\l\,\de_\l X^\m)\,\de\xx_\m
+Z\Ii\sA\sB\,\zzB\,\dezA~,
\\[6pt] \text{with}\quad
Z\Ii\sA\sB&\equiv
X^\l\,\de_\l Y'\Ii\sA\sB-X'^\l\,\de_\l Y\Ii\sA\sB
+Y'\Ii\sA\sC\,Y\Ii\sC\sB-Y\Ii\sA\sC\,Y'\Ii\sC\sB~.
\end{align*}
Moreover, if $Y$ and $Y'$ are Hermitian then $[Y,Y']$ is Hermitian.
\end{proposition}

Hence local Hermitian linear projectable vector fields
constitute a subsheaf of Lie algebras.

%--------------
\subsection{Hermitian vector fields and connections}
\label{ss:Hermitian vector fields and connections}

Let $c$ be a complex-linear connection of the bundle $\U\to\E$\,,
namely a section $\U\to\JU\subset\TE\ten{\E}\TU$,
with coordinate expression
$$c=\dx^\l\tn(\de\xx_\l+c\iIi\l\sA\sB\,\zzB\,\dezA)~,\quad
c\iIi\l\sA\sB:\E\to\CC~.$$
We say that $c$ is \emph{Hermitian} if $h$ is covariantly constant
relatively to $c$\,, \ie\ \hb{$\nabla[c]h=0$}\,.
It's immediate to check that, in coordinates,
the Hermiticity of $c$ means
that the matrices $\bigl(c\iIi\l\sA\sB\bigr)$\,, \hb{$\l=1,\dots,n$}\,,
are anti-Hermitian.

Any vector field $X:\E\to\TE$ can be lifted, via $c$\,,
to the linear projectable vector field $X\pint c:\U\to\TU$\,,
which has the coordinate expression
$$X\pint c=X^\l\,(\de\xx_\l+c\iIi\l\sA\sB\,\zzB\,\dezA)~,$$
namely $(X\pint c)\Ii\sA\sB=X^\l\,c\iIi\l\sA\sB$\,.
Clearly, $X\pint c$ is Hermitian if $ c$ is Hermitian.

Let $X,X':\E\to\TE$\,; then a straightforward calculation yields
$$[X\pint c,X'\pint c]=
(X ^\l\,\de_\l X'^\m-X'^\l\,\de_\l X ^\m)\,\de\xx_\m+
X ^\l\,X'^\m\,R[c]\iIi{\l\m}\sA\sB\,\zzB\,\dezA~,$$
where
$$R[c]\iIi{\l\m}\sA\sB=-\de_\l c\iIi\m\sA\sB+\de_\m c\iIi\l\sA\sB
+c\iIi\l\sA\sC\,c\iIi\m\sC\sB-c\iIi\m\sA\sC\,c\iIi\l\sC\sB$$
are the components of the curvature tensor
\hb{$R[c]:\E\to\weu2\TS\E\ten{\E}\End\U$}\,.
If $c$ is Hermitian, then $R[c]$ is an $\Lie$-valued 2-form.

A linear projection \hb{$\n[c]:\TU\to\VU$}
is associated with the  linear connection $c$\,.
Taking the natural identification \hb{$\VU\cong\U\cart{\E}\U$} into account,
the vertical projection of a linear projectable vector field
$Y:\U\to\TU$ can be seen as a section
$$\n[c]Y:\E\to\Lie\subset\End\U~,$$
with the coordinate expression
$$\n[c]Y=(Y\Ii\sA\sB-X^\l\, c\iIi\l\sA\sB)\,\zzB\tn\zeA~.$$
Thus $\n[c]Y$ is Hermitian if $c$ and $Y$ are Hermitian.

Conversely,
a Hermitian section $\chY:\E\to\End\U$
can be seen as a linear vertical vector field $\U\to\VU$\,;
for any given vector field $X:\E\to\TE$
we obtain the Hermitian linear projectable vector field
$$Y\!\!_c:=X\pint c+\chY:\U\to\TU~,$$
with coordinate expression
$$Y\!\!_c=
X^\l\,\de\xx_\l+(X^\l\, c\iIi\l\sA\sB+ \chY\Ii\sA\sB)\,
\zzB\,\dezA~.$$
Thus $Y\!\!_c$ is Hermitian if $c$ and $\chY$ are Hermitian.

Henceforth we assume, for simplicity,
that the considered connection $c$ is Hermitian
(though some of the forthcoming results also hold in a more general situation).
Let $\JC$ denote the sheaf of all pairs $(X,\chY)$
with $X:\E\to\TE$, $\chY:\E\to\Lie$\,,
and $\HC$ denote the sheaf of all Hermitian linear projectable
vector fields $Y:\U\to\TU$\,.
Then, clearly, the maps
$$\mathfrak{h}[c]:\HC\to\JC:Y\mapsto(X,\n[c]Y)~,\qquad
\mathfrak{j}[c]:\JC\to\HC:(X,\chY)\mapsto Y\!\!_c\equiv X\pint c+\chY~,$$
are inverse sheaf isomorphisms.

\begin{proposition}
Let $(X,\chY),(X',\chY')\in\JC$
and set $Y\!\!_c=X\pint c+\chY$,  $Y'\!\!\!_c=X'\pint c+\chY'$. Then
$$\n[c][Y\!\!_c\,, Y'\!\!\!_c\,]=
-R[c](X,X')+\na_X \chY'-\na_{X'}\chY+[\chY',\chY]~.$$
\end{proposition}\proof
A straightforward coordinate calculation yields
$$[Y\!\!_c\,, Y'\!\!\!_c\,]\Ii\sA\sB =[X,X']^\l\,c\iIi\l\sA\sB
-X^\l\,X'^\m\,R[c]\iIi{\l\m}\sA\sB
+X^\l\,\na_\l\chY'\Ii\sA\sB-X'^\l\,\na_\l\chY\Ii\sA\sB
+[\chY',\chY]\Ii\sA\sB~,$$
and applying the vertical projection $\n[c]$ amounts to subtracting
the term \hb{$[X,X']\pint c$}\,.
\qed

Now we are naturally led to introduce a Lie bracket on $\JC$,
in such a way that $\mathfrak{h}[c]$ and $\mathfrak{j}[c]$
turn out to be mutually inverse isomorphisms of Lie algebras.
Namely we set
\begin{align*}
[(X,\chY),(X',\chY')]_c&:=\Bigl(
[X,X']\,,~-R[c](X,X')+\na_X \chY'-\na_{X'} \chY+[\chY',\chY]\Bigr)\equiv
\\[6pt]
&\phantom:\equiv
\mathfrak{h}[c]\bigl([Y\!\!_c\,, Y'\!\!\!_c\,]\bigr)\equiv
\mathfrak{h}[c]\Bigl(
[\mathfrak{j}[c](X,\chY)\,,\mathfrak{j}[c](X',\chY')]\Bigr)~.
\end{align*}
Note that we have a different bracket and correspondence
for each Hermitian connection $c$\,.

%==============================================================================%
\section{Galileian spacetime and classical particle}
\label{s:Galileian spacetime and classical particle}
%--------------
\subsection{Galileian spacetime}
\label{ss:Galileian spacetime}

Classical spacetime is a basic ingredient of Covariant Quantum Mechanics.
Even in the Galileian case the theory can be formulated on a curved
background, describing a fixed gravitational field,
which is endowed with a rich structure.
We summarize some of the main notions which will be used here,
referring to other articles~\cite{JanMod02c}
for further developments and details.

The Galileian spacetime is assumed to be a fibered bundle
$$\tm:\E\to\T~,$$
where the spacetime manifold $\E$ is 4-dimensional
and the base manifold $\T$, the ``time'',
is an oriented 1-dimensional affine space.
The space of ``free vectors'' of $\T$  is written as $\RR\tn\TT$
where $\TT$ is a positive space\footnote{
A \emph{positive space}, or \emph{unit space},
is defined to be a semi-vector space $\UU$ on the semi-field $\RR^+$,
the action of $\RR^+$ on $\UU$ being free and transitive
(see~\cite{JMV08} for details).
The \emph{square root} $\UU^{1/2}$ of a unit space $\UU$,
is defined by the condition that $\UU^{1/2}\tn\UU^{1/2}$ be isomorphic to $\UU$.
More generally, any \emph{rational power} of a unit space
is defined up to isomorphism
(negative powers correspond to dual spaces).
The basic unit spaces commonly used are
the space $\LL$ of \emph{length units},
the space $\TT$ of \emph{time intervals}
and the space $\MM$ of \emph{masses}.
Coupling constants are elements in tensor products of positive spaces;
in particular, $\h\in\MM\tn\LL^2\tn\TT^{-1}$\,.} 
called the space of \emph{time intervals}.

The fibers of $\tm$ are assumed to be oriented as well
(for each $t\in\T$ the fiber over $t$\,,
namely the 3-dimensional submanifold $\E_t\equiv\tm^{-1}(t)$\,,
is the ``space at time $t$'').

A \emph{spacetime chart} is defined to be a fibered coordinate chart
$$\bigl(\xx^\l\bigr)\equiv\bigl(\xx^0,\xx^i\bigr):\E\to\RR\times\RR^3~.$$
Moreover we choose the coordinate $\xx^0$ such that
$u_0\equiv\bang{\dO\tm,\de\xx_0}\in\TT$ is a constant time interval;
then we write $\dO\tm=u_0\,\dx^0$\,.
Note that $u^0\equiv(u_0)^{-1}\in\TT^*$,
together with the choice of an ``origin'' $t_0\in\TT$\,,
determines a real-valued coordinate on $\T$.

Moreover we assume a \emph{spatial}, \emph{scaled} metric structure.
In general we say that a metric is scaled if it is valued in $\RR$
tensorialized by some positive space.
So we assume an $\LL^2$-scaled Riemannian metric
$g:\E\to\LL^2\tn\VS\E\ten{\E}\VS\E$\,,
where $\LL$ is the space of \emph{length units};
namely $g$ is a scaled metric on the fibers of $\E\to\T$,
since for each $t\in\T$ we have $(\VE)_t=\TO(\E_t)$\,.
In a spacetime chart we write
$$g=g_{ij}\,\check\dO\xx^i\tn\check\dO\xx^j~,
\quad g_{ij}:\E\to\RR\tn\LL^2~,~~i,j=1,2,3~,$$
where $\check\dO\xx^i$ indicates the \emph{vertical restriction}
of the 1-form $\dx^i:\E\to\TS\E$.

A connection $K$ of $\TE\to\E$ will be called a
\emph{spacetime connection} if it is linear, torsion-free,
and obeys the identities
\hb{$\nabla\dO\tm=0$}\,, \hb{$\nabla g=0$}\,,
\hb{$R_{ijhk}=R_{hkij}$}\,.

\remark~Differently from the Einstein case,
the spacelike metric $g$ by itself
does not fully characterize a distinguished connection of $\TE\to\E$\,.
Consider a linear spacetime connection $K$
fulfilling the condition \hb{$\nabla[K]\dO\tm=0$}\,.
In terms of its coordinate expression this means \hb{$K\iIi\l0\m=0$}\,,
namely $K$ is reducible to a connection $\check K$ of $\VE\to\E$\,;
in turn, this restricts to connections
on each spacelike manifold \hb{$\E_t\equiv\tm^{-1}(t)$}\,, \hb{$t\in\T$}
(the metricity condition \hb{$\nabla g=0$} implies that
these restrictions are just the torsion-free Riemannian connections
determined by $g$).
Then it's not difficult to see that the only part of $K$
which remains undetermined is the antisymmetric part
of the coefficients $K_{0jk}$\,,
namely the differences \hb{$K_{0jk}-K_{0kj}$}\,.
On turn, these can be characterized as the components
of closed 2-form $\Phi$ which depends
on the choice of an \emph{observer}
(\S\ref{ss:The phase space of a classical particle}).
This family of observer-dependent forms
is strictly related to the cosymplectic form $\Om$ (see below),
which is also closed.
The property of closeness turns out to be essential
in the quantum theory,
since it is necessary for the existence of the quantum connection.

\smallbreak
The classical particle mechanics
with fixed gravitational and electromagnetic background fields
can be formulated in the present context
by assuming that the gravitational field is described
by a given torsion-free metric spacetime connection $K^\nat$
fulfilling the above said conditions,
and that the electromagnetic field is described
by a closed scaled tensor field
$$F:\E\to(\MM\tn\LL)^{1/2}\tn\weu2\TS\E~.$$
With reference to a particle with mass $m$ and charge $q$\,,
$F$ can be merged with $K^\nat$ into a \emph{joined spacetime connection}
$K=K^\nat+K^{\mathfrak e}$
which obeys the same conditions as $K^\nat$\,.
In coordinates\footnote{
Namely one adds to $K^\nat$ the tensor
\hb{$K^{\mathfrak e}:\E\to\TS\E\tn\TE\tn\TS\E$}
obtained from $F\tn\dO\tm$
(times a suitable coupling constant)
by raising the second index and symmetrizing
relatively to the first and third indices
(the contravariant metric $g^\#$ can be naturally seen as a spacetime object,
not just a vertical object).} 
$$K\iIi ijk=K^\nat\iIi ijk~,\quad
K\iIi0jk=K^\nat\iIi 0jk+\frac q{2\,m}\,u_0\,F\Ii jk~,\quad
K\iIi0j0=K^\nat\iIi0j0+\frac qm\,u_0\,F\Ii j0~.$$

%--------------
\subsection{The phase space of a classical particle}
\label{ss:The phase space of a classical particle}

A \emph{particle motion} is a section \hb{$s:\T\to\E$}\,;
the first jet prolongation
\hb{$\jO s:\T\to\JE$}
is the particle's \emph{velocity}.
Hence we also call $\JE$ the \emph{phase space} for classical particle motions.
Since \hb{$\JE\to\E$} can be identified with a subbundle of
\hb{$\TT^*\tn\TE$}\,,
the velocity is a $\TT^*$-scaled vector.

A fibered spacetime chart $\bigl(\xx^0,\xx^i\bigr)$ determines
the fibered chart $\bigl(\xx^0,\xx^i,\xx^i_0\bigr)$ of $\JE$\,;
the velocity of $s$ has then the coordinate expression
\hb{$\jO s=u^0\,(\de\xx_0+s^i_0\,\de\xx_i)$}\,,
with \hb{$s^i_0\equiv\xx^i_0\comp\jO s=\de_0s^i$}\,.

An \emph{observer} is a connection of $\E\to\T$.
It can be seen as a section
\hb{$o:\E\to\JE$}\,,
hence it can also be seen as a $\TT^*$-scaled vector field on $\E$\,.
Its coordinate expression is $o=u^0\,(\de\xx_0+o^i_0\,\de\xx_i)$\,,
with $o^i_0:\E\to\RR$\,.
An observer can be seen as the field of velocity of a continuum,
whose integral motions are the ``horizontal sections'' of the connection $o$\,.
A spacetime chart is said to be \emph{adapted} to $o$ if $o^i_0=0$\,,
namely if the spatial coordinates $\xx^i$ are constant along
the integral motions of $o$\,.

The Galileian phase space has a rich structure,
whose detailed study lies outside the scope of this article
(\eg\ see~\cite{JanMod02c}).
As far as we are concerned, the two most important objects are
the 2nd order connection \hb{$\g:\JE\to\TT^*\tn\TO\JE$}\,,
induced by the joined spacetime connection $K$\,,
and the 2-form \hb{$\Om:\JE\to\weu2\TO^*\JE$}\,,
induced by $K$ and the ``rescaled'' metric\footnote{
Even in the classical theory Planck's constant
\hb{$\h\in\MM\tn\LL^2\tn\TT^{-1}$}
has a role (together with the particle's mass $m$)
in making the cosymplectic form $\Om$ a \emph{non-scaled} object.
We'll see that $\Om$ is an essential ingredient of the link
between the classical and the quantum theory.
Besides that, $\Om$ is needed at the classical level
in the study of symmetries~\cite{ModSalTol05}.} 
$\frac m\h\,g$
(with reference to a particle of mass $m$).
These objects can be intrinsically characterized in various ways;
their coordinate expressions are
\begin{align*}
\Om&=u^0\,\frac m\h\,g_{ij}\,
\bigl(\dx_0^i-(K\iIi\l ih\,\xx^h_0+K\iIi\l i0)\,\dx^\l\bigr)
\we(\dx^j-\xx_0^j\,\dx^0)~,
\\[6pt]
\g&=u^0\tn\bigl(\de\xx_0+\xx_0^i\,\de\xx_i
+(K\iIi0i0+2\,K\iIi0ij\,\xx_0^j+K\iIi hij\,\xx_0^h\,\xx_0^j)\,
\de\xx_i^0\bigr)~.
\end{align*}
Moreover one has \hb{$\Om=\Om^\nat+\frac{q}{2\,\h}\,F$}\,,
where $\Om^\nat$ is defined like $\Om$ but in terms of $K^\nat$ alone
rather than $K$.
One finds that the couple $(\dO\tm,\Om)$ is \emph{cosymplectic},
namely \hb{$\dO\Om=0$} and the 7-form
\hb{$\dO\tm\we\Om\we\Om\we\Om$} does not vanish.
For each observer $o$ the 2-form
\hb{$\Phi[o]\equiv 2\,o^*\Om$} is closed
(this is the same object,
describing that part of the spacetime connection
not determined by the spacelike metric $g$\,,
introduced in the remark in \S\ref{ss:Galileian spacetime}).
Similarly, $\g$ splits as \hb{$\g^\nat+\g^{\mathfrak e}$}\,,
where the gravitational part $\g^\nat$ is defined like $\g$
but in terms of $K^\nat$ rather than $K$,
and $\g^{\mathfrak e}$ turns out to be just the \emph{Lorentz force}.
The law of motion of a classical particle with mass $m$
and electric charge $q$\,,
in the given gravitational and electromagnetic fields,
can then be naturally written as\footnote{
If $\F\to\B$ is a fibered manifold and $\s:\B\to\F$ is a section,
then $\jO_k\s:\B\to\JO_k\F$ denotes the \emph{$k$-th jet prolongation}
of $\s$\,.
Moreover we use the shorthand $\jO\s\equiv\jO_1\s$\,.} 
\hb{$0=\nabla[\g]\jO s\equiv\jO_2s-\g\comp\jO s$}\,.

%--------------
\subsection{Special phase functions for a scalar particle}
\label{ss:Special phase functions for a scalar particle}

A certain Lie algebra of \emph{special functions} $\JE\to\RR$
can be intrinsically characterized in various ways.
The coordinate expression of a special function is of the type
$$f=f^0\,\frac{m\,u^0}{2\,\h}\,g_{ij}\,\xx_0^i\,\xx_0^j
+f^i\,\frac{m\,u^0}{\h}\,g_{ij}\,\xx_0^j+\br f~,$$
with $f^0,f^j,\br f:\E\to\RR$\,.
Note that the term $\br f$ depends on the coordinates.
In particular, for any observer $o$ we write $f[o]\equiv f\comp o$\,;
if the spacetime coordinates are \emph{adapted} to $o$\,,
so that $0=o_0^j\equiv\xx_0^j\comp o$\,, then $\br f=f[o]$\,.

Let $f,f'$ be functions of the above type, and define the bracket
$$\bbra{f,f'}:=\{f,f'\}+u_0\,(f^0\,\g).f'-u_0\,(f'^0\,\g).f~,$$
where $\{f,f'\}$ is the usual \emph{Poisson bracket}.
Then $\bbra{f,f'}:\JE\to\RR$ turns out to be a special function;
in coordinates adapted to the observer $o$ its expression is
\begin{align*}
&\bbra{f,f'}^\l=f^0\,\de_0f'^\l-f'^0\,\de_0f^\l
-f^h\,\de_h f'^\l+f'^h\,\de_h f^\l~,
\\[8pt]
&\bbra{f,f'}\!\breve{\phantom{A}}
=f^0\,\de_0\br f'-f'^0\,\de_0\br f
-f^h\,\de_h\br f'+f'^h\,\de_h\br f
-(f^0\,f'^h-f'^0\,f^h)\,\Phi_{0h}+f^h\,f'^k\,\,\Phi_{hk}~,
\end{align*}
with $\Phi_{\l\m}\equiv\Phi[o]_{\l\m}$\,.

Furthermore the above bracket turns out to fulfill the Jacobi identity.
Hence the local special functions constitute a sheaf of Lie algebras.

Any special function \hb{$f:\JE\to\RR$} determines,
by a geometric construction~\cite{JM06},
a vector field \hb{$X[f]:\E\to\TE$} whose coordinate expression is
\hb{$X[f]=f^0\,\de\xx_0-f^i\,\de\xx_i$}\,.
The map \hb{$f\mapsto X[f]$} turns out to be a morphism of Lie algebras
(with the usual Lie bracket of vector fields).

\remark~The Lie algebra of vector fields $X:\E\to\TE$ has the subalgebra
of all vector fields which are \emph{projectable} through $\E\to\T$.
This corresponds to the subalgebra of special functions $f$
such that \hbox{$\de_if^0=0$}\,,
that is \hbox{$f^0:\T\to\RR$}\,.
Actually, only special functions of this restricted type
turn out be physically meaningful~\cite{ModSalTol05}.
\remend

%--------------
\subsection[Special functions for a spin particle]{%
Phase space and special functions for a classical particle with spin}
\label{ss:Special functions for a spin particle}

We describe the motion of a classical particle with (classical) spin
as a section \hb{$u:\T\to\LL^*{\otimes}\VE$}.
This projects onto an ordinary particle motion \hb{$s:\T\to\E$}\,;
for each $t\in\T$
an ``intrinsic angular momentum'' \hb{$u(t)\in\LL^*\tn\VE_{s(t)}$}\,,
or \emph{spin}, is then associated to the particle.

The classical law of motion of a charged spin particle can then be formulated
in terms of spacetime connections,
using a further coupling constant
\hb{$\m\in\TT^{-1}\tn\LL^{3/2}\tn\MM^{-1/2}$}
(then $\mu{}u$ is the \emph{magnetic moment} of the particle).
For this purpose we introduce a further joined connection $K'$
which uses $2\,\m$ rather than $q/m$ as the coupling constant
between $K^\natural$ and $F$ (\S\ref{ss:Galileian spacetime}).
Namely we set $K':=K^\natural+K^{\mathfrak{m}}$,
where $K^{\mathfrak{m}}$ is obtained from $2\,\m\,F\tn\dO\tm$
by performing on it the same algebraic operations
which yield $K^{\mathfrak{e}}$.
In coordinates we have
$$K'\iIi ijk=K^\nat\iIi ijk~,\quad
K'\iIi0jk=K^\nat\iIi 0jk+\m\,u_0\,F\Ii jk~,\quad
K'\iIi0j0=K^\nat\iIi0j0+2\,\m\,u_0\,F\Ii j0~.$$
Then the equation of motion for $u$ can be formulated as
$$0=\nabla[\g,\g']u'\equiv\jO u'-(\g,\g')\comp u'~,$$
where $\g'$ is the second order connection
(\S\ref{ss:The phase space of a classical particle})
associated with $K'$ and we used the shorthand
\hb{$u'\equiv(\jO s,u):\E\to\JE\cart{\E}\LL^*\tn\VE$}.
Equivalently, the same equation can also be written as
\hb{$\na_{\jO s}u-\m\,u{\times}B=0$}\,,
where $B:\E\to\LL^{-5/2}\tn\MM^{1/2}\tn\VE$
is the magnetic field\footnote{
\label{foot:magneticfield}
In the Galileian context this is observer-independent,
being defined as
\hb{$B:=\oh\,{*}\,\check F$}
where $\check F$ is the vertical restriction of $F$
\cite{CJM95,JadMod94}.} 
associated with $F$.
One then finds that $s(t)$ coincides with the usual motion
of a scalar charged particle,
while the spin only interacts with the magnetic field.

The main point of interest here is that one is led
to identify the phase space of a classical particle with spin
with the first jet prolongation $\JO(\LL^*\tn\VE)$\,.
However, a smaller phase space turns out to suit our purposes,
namely we take\footnote{
We recall that there is a natural isomorphism \hb{$\JO\VE\cong\VO\JE$}\,.
It is easy to see that we have an analogous isomorphism
\hb{$\JO(\LL^*\tn\VE)\cong\LL^*\tn\VO\JE$}\,.
Hence this extended phase space has two distinct
natural projections onto $\JE$ and $\LL^*\tn\VE$\,.
} 
\hb{$\JE\cart{\E}(\LL^*{\otimes}\VE)$}\,.
This choice can be justified by various arguments,
including the observation that
one does not measure a ``spin speed'' in quantum mechanics;
a point is even more relevant in the present context:
a sound confirmation will be found
in the relation between special functions and Hermitian vector fields.

So we consider special functions of the type
$$f+\phi\equiv (f\comp\pr1+\phi\comp\pr2)
:\JE\cart{\E}(\LL^*{\otimes}\VE)\to\RR~,$$
where \hb{$f:\JE\to\RR$} is assumed to be a special function in the sense
of the Covariant Quantum Mechanics of a scalar
particle (\S\ref{ss:Special phase functions for a scalar particle}).
As for $\phi:\LL^*\tn\VE\to\RR$
we simply assume it to be a linear map over $\E$,
namely a section \hb{$\E\to\LL\tn\VS\E$}\,.
%In coordinates
%$$\phi=\phi_i\,\Pau^i~,\quad
%\Pau^i:=\Pau_*\xi^i=\cev\Pau{}^i_j\,\dx^j~.$$

%----------
\subsection{Lie bracket of extended special functions}
\label{ss:Lie bracket of extended special functions}

The next step consists in extending the Lie algebra of special functions
for the scalar case to a larger Lie algebra of functions of the above type.
We begin by doing a preliminary construction
based on the curvature tensor of the connection $\check K$ of $\VE\to\E$
induced by $K$.
This tensor is a section\footnote{
$\End(\VE)=\VE\ten{\E}\VS\E\cong(\LL^*\tn\VE)\ten{\E}(\LL\tn\VS\E)
=\End(\LL^*\tn\VE)$\,.} 
$$\check R:\E\to\weu2\TS\E\ten{\E}\End(\VE)\cong
\weu2\TS\E\ten{\E}\End(\LL^*\tn\VE)~.$$
Now there is a natural fibered isomorphism
\hb{$\tri:\End(\VE)\to\LL\tn\VS\E$}
over $\E$, obtained by index-raising the second factor
in \hb{$\End\E\cong\VE\ten{\E}\VS\E$}
via the spacelike metric $g$ and then contracting
by the spacelike volume form associated with $g$\,.
Then we get a map
$$\tri:\End(\VE)\to\LL\tn\VS\E~.$$
We define
$$\r:=\check R\pint \tri:\E\to\LL\tn\weu2\TS\E\ten{\E}\VS\E~.$$
If $\bigl(\ee_i\bigr)$ is an orthonormal frame of $\LL^*\tn\VO\E$\,,
and
\hb{$\check R=\check R\iIi{\l\m}ij\,\dx^\l\we\dx^\m\tn\ee_i\tn \check\ee^j$}\,,
then we find
$$\r=\r_{\l\m k}\,\dx^\l\we\dx^\m\tn\check\ee^k~,\qquad
\r_{\l\m k}\equiv \check R\iI{\l\m}{ij}\,\e_{ijk}\,.$$

\remark~Actually, the curvature tensor $\check R$ is a section
\hb{$\check R:\E\to\weu2\TS\E\ten{\E}\Aie$}\,,
where
$$\Aie\subset\VE\ten{\E}\VS\E\equiv(\LL^*\tn\VE)\ten{\E}(\LL\tn\VS\E)$$
is the subbundle of all $g$-antisymmetric endomorphisms.
It is easy to check that $\Aie$ is closed
with respect to the ordinary commutator,
so that its fibers are equipped with a Lie algebra structure.\footnote{
This is essentially the Lie algebra of $\SO(3)$\,.} 
On the other hand, the fibers of $\LL^*\tn\VE$ are
equipped with the Lie algebra structure given by the cross product.
Then it can be checked that $-\oh\,\tri$
is  a Lie-algebra isomorphism.\rembox

Now for \hb{$f{+}\phi,f'{+}\phi':\JE\cart{\E}(\LL^*{\otimes}\VE)\to\RR$} we set
\begin{align*}
&\bbra{f{+}\phi,f'{+}\phi'}=
\bbra{f,f'}+\bbra{f{+}\phi,f'{+}\phi'}\!\check{\phantom{A}}
\\[6pt]
\text{with}\quad
&\bbra{f{+}\phi,f'{+}\phi'}\!\check{\phantom{A}}
:=-\r(X[f],X[f'])+\na_{X[f]}\phi'-\na_{X[f']}\phi+\phi'\,{\times}\,\phi~;
\end{align*}
here
$\bbra{f,f'}$ is the bracket of special functions for a scalar particle
and $X[f]:\E\to\TE$ is the vector field determined by $f$\,,
whose coordinate expression is \hb{$X[f]=X^\l\,\de\xx_\l$} with
\hb{$X^0=f^0$}\,, \hb{$X^i=-f^i$}
(\S\ref{ss:Special phase functions for a scalar particle}).
Using an orthonormal frame $\bigl(\ee_i\bigr)$ of \hb{$\LL^*\tn\VE$}\,,
the coordinate expression of the last part in the above bracket is
\begin{align*}
& \bbra{(f{+}\phi),(f'{+}\phi')}\!\check{\phantom{A}}_{\!\!\!\!k}=
\\[6pt]
&\qquad
=-\r_{\l\m\,k}\,X^\l\,X'^\m
+X^\l\,(\de_\l\phi'_k+\td K\iIi\l jk\,\phi'_j)
-X^\l\,(\de_\l\phi'_k+\td K\iIi\l jk\,\phi'_j)
-\phi^i\,\phi'^j\,\e_{ijk}~,
\\[8pt]
& X^0=f^0~,~X^i=-f^i~,
\end{align*}
where $\td K\iIi\l jk$ are the coefficients of $\check K$
in the frame $\bigl(\ee_i\bigr)$\,.

The above bracket turns out to obey the Jacobi identity,
hence it determines a structure of sheaf of Lie algebras
in the sheaf of all local phase functions for a spin particle.
This property is a consequence of the correspondence
between special functions and Hermitian vetor fields,
which will be proved
in~\S\ref{ss:Hermitian vector fields and special functions}.
However, a direct check can be instructive.

%-----------
\subsection{Checking the Jacobi identity}
\label{ss:Checking the Jacobi identity}

The Jacobi identity for the bracket $\bbra{f,f'}$
of the special functions for the scalar case has already been
established in previous works~\cite{ModSalTol05}.
As for the remaining part, we get
\begin{align*}
&\bbra{(f_1{+}\phi_1),
\bbra{(f_2{+}\phi_2),(f_3{+}\phi_3)}}\!\check{\phantom{A}}=
\\[10pt]
&~=-\r(X_1,[X_2,X_3])-(\na_{X_1}\r)(X_2,X_3)
-\r(\na_{X_1}X_2,X_3)-\r(X_2,\na_{X_1}X_3)+{}
\\[6pt]
&\qquad +\na_{X_1}\na_{X_2}\phi_3-\na_{X_1}\na_{X_3}\phi_2
-\na_{[X_2,X_3]}\phi_1-\r(X_2,X_3)\,{\times}\,\phi_1+{}
\\[6pt]
&\qquad +\na_{X_1}(\phi_3\,{\times}\,\phi_2)
+(\na_{X_2}\phi_3-\na_{X_3}\phi_2)\,{\times}\,\phi_1
+(\phi_3\,{\times}\,\phi_2)\,{\times}\,\phi_1~.
\end{align*}

We must check that the sum of the above expression
over the cyclic permutations of the set $(1,2,3)$ vanishes.
Now we observe that this is a sum of four parts which must vanish independently,
containing a different number of $\phi$ factors.

Consider the part not containing the $\phi$ fields
(namely the first line in right-hand side
of the above expression).
Using the assumption that the spacetime connection
be torsion-free and summing up the above expression
over the cyclic permutations of the set $(1,2,3)$ one sees that,
because of the antisymmetry of $\r$\,,
all terms containing the covariant derivatives of the fields $X_i$
eventually cancel out, so that we are left with
$$-(\na_{X_1}\r)(X_2,X_3)-(\na_{X_2}\r)(X_3,X_1)-(\na_{X_3}\r)(X_1,X_2)=0~,$$
vanishing because of the \emph{Bianchi identities}\footnote{
The vanishing-torsion condition reads
\hb{$\na_{X_1}X_2-\na_{X_2}X_1=[X_1,X_2]$}\,, and the like.
However the constructions and results of this section do not actually
depend on this condition.
If we allow for a non-vanishing torsion then checking the Jacobi identity
is somewhat more complicated
and uses the following generalized form of the Bianchi identities:
$$\na_\n\r\iI{\l\m}i+\na_\l\r\iI{\m\n}i+\na_\m\r\iI{\n\l}i=
\r\iI{\n\s}i\,T^\s_{\l\m}+\r\iI{\l\s}i\,T^\r_{\m\n}+
\r\iI{\m\s}i\,T^\s_{\n\l}~.$$
Similar identities can be shown to hold for the curvature tensor
of any linear connection of a vector bundle,
provided that one also has a linear connection of the base.} 
$\na_\n\r\iI{\l\m}i+\na_\l\r\iI{\m\n}i+\na_\m\r\iI{\n\l}i=0$\,.

Next we consider the second line in the above expression of
\hb{
$\bbra{(f_1,\phi_1),\bbra{(f_2,\phi_2),(f_3,\phi_3)}}\!\check{\phantom{A}}$}.
After summing over cyclic permutations,
and taking into account
the definition of the curvature tensor, we get the sum
\begin{align*}
&\ost R(X_1,X_2)\phi_3-\r(X_1,X_2)\,{\times}\,\phi_3+
\ost R(X_2,X_3)\phi_1-\r(X_2,X_3)\,{\times}\,\phi_1+{}
\\[6pt]
&\qquad{}+\ost R(X_3,X_1)\phi_2-\r(X_3,X_1)\,{\times}\,\phi_2~,
\end{align*}
where $\ost{R}$
is the curvature tensor of the connection of \hb{$\VS\E\to\E$}
induced by $\check K$\,.
The latter expression vanishes since
for any sections \hb{$X,X':\E\to\TE$} and \hb{$\phi:\E\to\VS\E$}
we have
\begin{align*}
\bigl(\ost{R}(X,X')\phi\bigr){}_j&=-X^\l\,X'^\m\,R\iIi{\l\m}kj\,\phi_k
=-X^\l\,X'^\m\,R\iI{\l\m}i\,\e\iI{ij}k\,\phi_k
=\bigl(\r(X,X')\,{\times}\,\phi\bigr){}_j~.
\end{align*}

As for the part of the double bracket whose terms
contain two fields $\phi_i$\,,
after summing over the cyclic permutations we get
an expression which clearly vanishes
because of the antisymmetry of the cross product.
Finally, the sum over the cyclic permutations of
\hb{$(\phi_3\,{\times}\,\phi_2)\,{\times}\,\phi_1$}
vanishes because of the Jacobi property obeyed by the cross product.

%==============================================================================%
\section{Spinors for the Galileian setting}
\label{s:Spinors for the Galileian setting}
%--------------
\subsection{Pauli spinors} \label{ss:Pauli spinors}

Consider a complex 2-dimensional vector space $\U$
endowed with a positive Hermitian metric $h$\,.
The Lie algebra $\Lie_0\subset\End\U$
of all traceless anti-Hermitian endomorphisms of $\U$ is (real) 3-dimensional,
and is naturally endowed with the Euclidean metric
$$\td g:\Lie_0\times\Lie_0\to\RR:(A,B)\mapsto-2\,\Tr(A\comp B)~.$$
If $\bigl(\xi_i\bigr)$ is any
$\td g$-orthonormal basis of $\Lie_0$\,,
then we find
$$[\xi_i\,,\,\xi_j]=\pm\e\iI{ij}k\,\xi_k\equiv\pm\e_{ijh}\,\d^{hk}\,\xi_k~.$$
We'll use orthonormal bases so oriented that the plus sign
holds in the above formula
(namely the coefficients $\e\iI{ij}k$ are the `structure constants'
of $\Lie_0$).

\remark~
Let $\bigl(\zeA\bigr)$\,, ${\scriptstyle{A}}=1,2$\,,
be an $h$-orthonormal basis of $\U$,
and let $\bigl(\zzA\bigr)$ be its dual basis.
The above statements are easily checked
by means of the associated \emph{Pauli basis},
that is the positively oriented
$\td g$-orthonormal basis $\bigl(\xi_i\bigr)$ of $\Lie_0$ defined as
\begin{align*}
&\xi_i\equiv-\ih\,\s_i\equiv-\ih\,\s\iIi i\sA\sB\,\zeA\tn\zzB~,\quad i=1,2,3,
\\[6pt]
& \bigl(\s_1\bigr)=
\le(\begin{smallmatrix}~0&~\hm1~\\~1&~\hm0~\end{smallmatrix}\ri)~,\qquad
\bigl(\s_2\bigr)=
\le(\begin{smallmatrix}~0&~-\iO~\\~\iO&~\hm0~\end{smallmatrix}\ri)~,\qquad
\bigl(\s_3\bigr)=
\le(\begin{smallmatrix}~1&~\hm0~\\~0&~-1~\end{smallmatrix}\ri)~.
\end{align*}
Conversely, any other orthonormal basis $\bigl(\xi'_i\bigr)$
with the same orientation
is a Pauli basis for some $h$-orthonormal basis $\bigl(\ze'_\sA\bigr)$ of $\U$,
which is unique up to sign
(this is essentially
the \hb{2-to-1} covering \hb{$\SU(2)\to\SO(3)$}\,).\rembox

Let now $\U\to\E$ be a complex vector bundle over classical spacetime,
with 2-dimensional fibers smoothly endowed with a Hermitian metric $h$\,.
The fibers of $\Lie_0\to\E$ are then smoothly orientable,
and we have a distinguished orientation determined by the condition
that Pauli frames be positively oriented.

We'll assume a \emph{Pauli map},
that is an orientation-preserving fibered isometry
$$\Pau:\LL^*\tn\VE\to\Lie_0$$
over $\E$.
Obviously $\Pau$ determines a bijection
among positively oriented
orthonormal frames $\bigl(\ee_i\bigr)$ of $\LL^*\tn\VE$
and $\bigl(\xi_i\bigr)$ of $\Lie_0$\,.
Since $\ee_i\,{\times}\,\ee_j=\e_{ij}{}^k\,\ee_k$\,,
we see that $\Pau$ is an isomorphism of Lie algebras,
the Lie algebra of $\LL^*\tn\VE$ being given by the ordinary cross product.

%--------------
\subsection{Spin connection} \label{ss:Spin connection}

A \emph{spin connection} is a linear connection $C$ of $\U\to\E$.
It has the coordinate expression
\hb{$C=\dx^\l\tn(\de\xx_\l+C\iIi\l\sA\sB\,\zzB\,\dezA)$}\,,
with \hb{$C\iIi\l\sA\sB:\E\to\CC$}\,.
We set $\xi_0\equiv\iO\,\s_0\equiv\iO\,\id$\,, and write
$$C\iIi\l\sA\sB\equiv C_\l^\m\,\xi\iIi\m\sA\sB
=\iO\,C_\l^0\,\dAB-\ih\,C_\l^j\,\s\iIi j\sA\sB~.$$
In the present Galileian context we require that $C$ be Hermitian,
\ie\ $\nabla[C]h=0$\,.
This amounts to say that the coefficients $C_\l^\m$ be \emph{real},
namely that the $2\,{\times}\,2$ matrices
$\bigl(C\iIi\l\sA\sB\bigr)$ be anti-Hermitian,
$\l=0,1,2,3$.

The connection of $\End\U\cong\U\ten{\E}\Ul\to\E$ induced by $C$
turns out to be reducible to a connection $\td C$ of $\Lie_0\to\E$\,,
whose coefficients in the frame $\bigl(\xi_i\bigr)$ are
$$\td C\iIi\l kj=C_\l^i\,\e\iI{ij}k~.$$
Then it's easy to check that $\td C$ is \emph{metric},
\ie\ $\nabla[\td C]\td g=0$\,.

A similar relation holds between the curvature tensors
$$R[C]:\E\to\weu2\E\tn\U\tn\Ul~,\qquad
\td R\equiv R[\td C]:\E\to\weu2\E\tn\Lie_0\tn\Lie_0^*~,$$
of $C$ and $\td C$\,.
Actually $R[C]$ turns out to be a 2-form valued into the anti-Hermitian
endomorphisms of $\U$\,,
so we write its coordinate expression as
$R[C]\iIi{\l\m}\sA\sB=R\iI{\l\m}\n\,\xi_\n$\,,
with $R\iI{\l\m}\n:\E\to\RR$\,.
By a simple calculation we then find
$$R\iI{\l\m}0=-\de_\l C_\m^0+\de_\m C_\l^0~,\quad
R\iI{\l\m}k=-\de_\l C_\m^k+\de_\m C_\l^k
+C_\l^i\,C_\m^j\,\e\iI{ij}k~.$$
Moreover (by a slightly more complicate calculation) we also find
$$\td R\iIi{\l\m}kj=R\iI{\l\m}i\,\,\e\iI{ij}k~.$$

\remark~The fibers of \hb{$\weu2\U\to\E$} are naturally endowed
with a Hermitian structure induced by $h$\,;
if $C$ preserves $h$\,,
then the induced connection $\hat C$ of \hb{$\weu2\U\to\E$}
preserves $\hat h$\,.
The expression of $\hat C$
in the (orthonormal) frame \hb{$\ze_1\we\ze_2$} is
$$\hat C_\l=-C\iIi\l\sA\sA=-2\,\iO\,C_\l^0~.$$
Now we observe that,
while a spin connection $C$ preserving $h$ determines a unique
metric linear connection $\td C$ of \hb{$\Lie_0\to\E$}\,,
the converse is not true:
$\td C$ only determines $C$ \emph{in part}.
More precisely, $\td C$ does not determine the coefficients $C_\l^0$\,,
namely it does not determine the \emph{trace}
$C\iIi\l\sA\sA$ of the coefficients of $C$ in an $h$-orthonormal frame.
In other terms, $\td C$ does not determine
the Hermitian connection\footnote{
$C\iIi\l\sA\sA$ in only known to be imaginary,
because \hb{$\nabla h=0$}\,.} 
$\hat C$\,.
\remend\smallbreak

If we assume a given spacetime connection $K^\nat$,
representing the gravitational field,
then the assigned Pauli map $\Pau$
determines a metric connection $\td C$ of $\Lie_0\to\E$
(we get $\nabla\Pau=0$)\,.
On turn, this determines the trace-free part of a Hermitian spin connection,
that is the coefficients $C_\l^i$ 
(\hb{$\l=0,1,2,3$}, \hb{$i=1,2,3$})\,.
Moreover, note that
$$R\iI{\l\m}i=\r\iI{\l\m}i~,$$
where the $\r\iI{\l\m}i$ are the components of the field
\hb{$\r:\E\to\LL\tn\weu2\TS\E\ten{\E}\VS\E$}
introduced in~\S\ref{ss:Lie bracket of extended special functions}.

Recalling the above remark, we see that the chosen spacetime connection,
together with the Pauli map,
determines an $h$-preserving spin connection
up to the induced $\hat h$-preserving
linear connection of \hb{$\weu2\U\to\E$}\,.

\remark~Conversely, we could assume a spin connection $C$
as a primary datum and require that $K^\nat$ restricts
to $\check K^\nat$ determined by $C$ through $\Pau$\,.
In general, such a spacetime connection\footnote{
In any case, $K$ is not fully determined by $C$\,.} 
would have non-vanishing torsion
(unless further suitable conditions on $C$ are required).
Since in the present work we deal with a fixed spacetime background,
maintaining a torsion-free spacetime connection
as a primary geometric datum seems reasonable.
Taking a further step we could study the interaction
between spin and gravitation in the Galileian context,
possibly finding that spin is a source for torsion
like in the theory of Einstein-Dirac coupled fields~\cite{C98,C00b,C07}.\rembox

%--------------
\subsection{Relation to 2-spinors in Einstein spacetime}
\label{ss:Relation to 2-spinors in Einstein spacetime}

We refer to the treatment of 2-spinors and spacetime
that was exposed in previous articles~\cite{C98,C00b,C07}.
While that exposition has some original aspects
which clarify the relations among the fundamental geometric data
of Einstein-Cartan-Maxwell-Dirac field theory,
it can be shown to be essentially equivalent to the standard theory.

Assume that the 2-spinor space $\U$ is only equipped
with a Hermitian structure of $\weu2\U$,
while no Hermitian metric $h$ on $\U$ itself is assigned.
Then a normalized ``complex symplectic form'' $\e\in\weu2\Ul$
is determined up to a phase factor,
so $\e\tn\be$ is unique and is seen to restrict to a natural Lorentz metric
on the Hermitian subspace of $\H\subset\U\tn\Uc$;
this real vector space is also naturally endowed with a Clifford map
$\g:\H\to\End(\W)$,
where $\W:=\U\oplus\Ua$ can be identified with the space
of \emph{Dirac spinors}.
A Hermitian metric $h$ can be identified with a timelike element of $\H^*$,
namely its assignment is equivalent to that of an \emph{observer},
and determines a linear splitting \hb{$\H=\H^\spar\oplus\H^\sbot$}
(mutually orthogonal `time' and `space').
Moreover $h$ also determines
(via `index lowering' and multiplication by $\iO$)
an isomorphism $\H\leftrightarrow\Lie$\,;
more precisely, $\H^\sbot$ is associated with the traceless subspace
$\Lie_0\subset\Lie$,
while $\H^\spar$ is associated with the subspace $\iO\,\RR\,\Id{\U}$
generated by the identity.

Let now \hb{$\U\to\E$} be a bundle, over the 4-dimensional real manifold $\E$,
with fibers as above.
A fibered isomorphism \hb{$\Th:\TO\E\to\LL\tn\H$} over $\E$
(a \emph{tethrad}, or \emph{soldering form})
transforms the Lorentz metric of $\H$ to a scaled Lorentz metric of $\E$.
A linear connection $C$ of $\U\to\E$ yields a metric connection of $\H\to\E$
and, via the condition $\nabla\Th=0$\,,
a metric connection of \hb{$\TO\E\to\E$}.
A Hermitan structure $h$ of $\U$ the yields, via $\Th$\,, an observer on $\E$
(a timelike unit vector field $\E\to\LL^*\tn\TO\E$).

Summarizing, we can say that the general relativistic situation
with a chosen observer is somewhat similar to the Galileian situation.

%==============================================================================%
\section{Quantum setting}\label{s:Quantum setting}

%--------------
\subsection{Quantum setting for a scalar particle}
\label{ss:Quantum setting for a scalar particle}

We recall from~\S\ref{ss:Galileian spacetime}
that in Galileian spacetime $\E\to\T$
one assumes the following basic structures
for the classical mechanics of a particle with no internal structure:
the spacelike metric $g$\,,
the gravitational connection $K^\nat$ and
the electromagnetic field $F$.
The mass $m$ and the charge $q$ of the particle
allow the last two objects to be assembled
into the joined spacetime connection $K$,
which in turn yields the cosympletic 2-form $\Om$ on $\JE$.

The above recalled classical objects
are required for the Covariant Quantum Mechanics
of a particle of the said type;
moreover one takes a complex vector bundle \hb{$\Q\to\E$}
with 1-dimensional fibers,
equipped with a Hermitian metric $h_{{\sst\Q}}$
and a \emph{quantum connection};
this is defined to be a Hermitian linear connection $\Ch$
of the \emph{phase quantum bundle}
\hb{$Q^\up\equiv\JE\cart{\E}\Q\to\JE$}\,,
fulfilling the two following conditions:
it is a \emph{universal} connection,
and its curvature $R[\Ch]$ is proportional
to the cosymplectic form $\Om$
(which contains the Planck constant)
through the relation
$$R[\Ch]=-2\,\iO\,\Om\tn\ii~,$$
where
$\ii:\Q\to\VO\Q\cong\Q\cart{\E}\Q:z\mapsto(z,z)$
is the \emph{Liouville vector field}
(in coordinates \hb{$\ii=\zz\,\de\zz$}\,).

\remark~In Hermitian spaces we use normalized frames.
Here $\zz$ denotes the dual frame (linear fiber coordinates)
of a normalized frame $\ze:\E\to\Q$\,.~\remend

The condition that $\Ch$ be universal
can be expressed in coordinates\footnote{
The general coordinate expression of a linear connection of $\Q^\up\to\JE$
is $\Ch=\dx^\l\tn(\de\xx_\l+\iO\,\Ch_\l\,\zz\,\de\zz)
+\dx_0^i\tn(\de\xx^0_i+\iO\,\Ch_i^0\,\zz\,\de\zz)$\,.} 
as $\Ch_i^0=0$\,,
and implies that, for each observer $o$\,, the pull-back
\hb{$\Ch[o]:=o^*\Ch$} is a Hermitian
linear connection of \hb{$\Q\to\E$}\,;
a certain transformation then relates the
connections determined by any two observers.

For each observer, \hb{$\Phi[o]\equiv\iO\,\Tr R[\Ch[o]]$}
turns out to be a closed 2-form \hb{$\E\to\weu2\TS\E$}
(this is the same object that was already considered
in~\S\ref{ss:The phase space of a classical particle}).
Choose a spacetime chart adapted to $o$
and a local \emph{normalized} frame of \hb{$\Q\to\E$}\,;
then $\Ch$ turns out to have the coordinate expression
$$\Ch=\dx^\l\tn\de\xx_\l+\dx_0^i\tn\de\xx^0_i
+\iO\,\Ch_\l\,\dx^\l\tn\ii~,$$
with
$$\Ch_0=-\frac{m\,u^0}{2\,\h}\,g_{ij}\,\xx_0^i\,\xx_0^j+A_0~,
\qquad
\Ch_i=\frac{m\,u^0}{\h}\,g_{ij}\,\xx_0^j+A_i~,$$
where $A[o]=A_\l\,\dx^\l$ is a distinguished potential of $\Phi[o]$\,,
determined by $\Ch$ and the chosen (normalized) quantum frame.
Then \hb{$\Hcal_0\equiv-\Ch_0$}
and \hb{$\Pcal_i\equiv\Ch_i$} are the
\emph{classical Hamiltonian and momentum} of the particle.

The question of the existence of a quantum connection,
and of how many quantum connections exist,
is essentially of cohomological nature~\cite{Vit99}.

%----------
\subsection{Quantum connection for a spin particle}
\label{ss:Quantum connection for a spin particle}

Let $\U\to\E$ be the bundle of Pauli spinors
(\S\ref{s:Spinors for the Galileian setting}),
endowed with a Hermitian metric $h$
and related to the spacetime geometry by the Pauli map $\Pau$\,.
We extend the quantum setting
sketched in~\S\ref{ss:Quantum setting for a scalar particle}
by introducing the \emph{Pauli phase quantum bundle}
\hb{$\U^\up\equiv\JE\cart{\E}\U\to\JE$}\,.

In order to extend the scalar case construction
we must postulate a suitable connection on \hb{$\U^\up\to\JE$}\,,
which can be obtained in a natural way by the following argument.
First, we write the bundle of Pauli spinors as a tensor product
$$\U=\Q\ten{\E}\U'~,$$
that is, equivalently, we set \hb{$\U':=\Q^\lin\tn\U$}\,.
Note that the fibers of \hb{$\U'\to\E$} are naturally endowed
with the Hermitian structure induced by $h_{{\sst\Q}}$ and $h$\,,
and that we have the obvious isomorphism $\End\U'\cong\End\U$\,.
We also have natural isomorphisms
\hb{$\Lie\cong\Lie'$} and \hb{$\Lie_0\cong\Lie'_0$}\,,
between the bundles of all anti-Hermitian endomorphisms
and between the bundles of all trace-free anti-Hermitian endomorphisms
of $\U$ and $\U'$\,.
Hence the connection $\td C$ of \hb{$\Lie_0\to\E$}
(determined by the spacetime connection $K$
and the Pauli map \hbox{$\Pau:\LL^*\tn\VE\to\Lie_0$}, 
\S\ref{ss:Spin connection})
yields Hermitian spin connections
of \hb{$\U\to\E$} and \hb{$\U'\to\E$}
up to the induced connections of \hb{$\weu2\U\to\E$} and \hb{$\weu2\U'\to\E$}
(\S\ref{ss:Spin connection}).
In mutually proportional orthonormal frames of $\U$ and $\U'$\,,
the trace-free components of these two spin connections are identical.

Now we \emph{choose} a suitable spin connection $C$ of \hb{$\U\to\E$}\,,
compatible with $\td C$\,.
The simplest natural way to restrict the choice of $C$
consists of assuming that
\emph{the induced connection of \hb{$\weu2\U'\to\E$} is flat}
(namely its curvature tensor vanishes---this implies
that certain topological conditions must be fulfilled).
Then it's not difficult to see that, locally,
one can find orthonormal spin frames such that
the coefficients $C\!\iIi\l\sA\sB$ are trace-free
(\ie\ the induced frame of \hb{$\weu2\U'\to\E$} is covariantly constant).

Eventually, we take the connection $\Ch\tn C$ of
$$\U^\up=\Q^\up\ten{\E}\U'\to\JE$$
determined by the spin connection $C$
and by the quantum connection $\Ch$ of the scalar case.
Its coordinate expression is
$$\Ch\tn C'=\dx^\l\tn\de\xx_\l+\dx_0^i\tn\de\xx^0_i
+\dx^\l\tn(\iO\,\Ch_\l\,\d\Ii\sA\sB+C\iIi\l\sA\sB)\,
\zzB\,\dezA~,$$
with
$C\iIi\l\sA\sB=C_\l^i\,\xi\iIi i\sA\sB$\,, $i=1,2,3$\,,
$C_\l^i:\E\to\RR$\,.

Thus $\Ch\tn C$ can be seen as a universal connection of
$\U^\up\equiv\JE\cart{\E}\U\to\JE$\,,
namely a certain family of connections $\Ch[o]\tn C$ of $\U\to\E$
parametrized by the observers.
The components of its curvature tensor can be similarly written as
\begin{align*}
R[\Ch\tn C]\iIi{\l\m}\sA\sB\equiv
R[\Ch\tn C]\iI{\l\m}\n\,\xi\iIi\n\sA\sB
&=R[\Ch]_{\l\m}\,\dAB+R[C]\iIi{\l\m}\sA\sB=
\\[6pt]
&=-2\,\iO\,\Om_{\l\m}\,\dAB+\r\iI{\l\m}k\,\xi\iIi k\sA\sB~.
\end{align*}

\remark~If the complex line bundle $\weu2\U$ has a \emph{square root},
then this could be identified with $\Q$\,;
in other terms we write \hb{$\Q\tn\Q\cong\weu2\U$}\,,
and identify $\U'$ with \hb{$\Q^\lin\ten{\E}\U$}\,.\rembox

%----------
\subsection{Hermitian vector fields and special functions}
\label{ss:Hermitian vector fields and special functions}

In this section we'll see that there is a one-to-one
correspondence among the special functions
\hb{$\JE\cart{\E}(\LL^*\tn\VE)\to\RR$} considered
in~\S\ref{ss:Special functions for a spin particle}
and Hermitian vector fields \hb{$\U\to\TU$}.
Moreover that correspondence will turn out to be an isomorphism
of Lie-algebra sheaves.
The correspondence can be established in two steps,
via an intermediate passage
which depends on the choice of an observer.
The final result, however, is independent of the observer
(provided that one uses the same observer in both steps).

We begin by recalling the results stated
in~\S\ref{ss:Hermitian vector fields and connections},
which enable us to use a Hermitian linear connection of $\U\to\E$
in order to characterize any Hermitian vector field.
In the present case we use the connection $\Ch[o]\tn C$
determined by the choice of an observer $o$\,.
\begin{theorem}\label{fieldcharacterization}
Let $Y:\U\to\TU$ be a Hermitian linear projectable vector field
and $o:\E\to\JE$ an observer.
Then the connection $\Ch[o]\tn C$ of $\U\to\E$
determines the pair $\bigl(X,\chY[o] \bigr)$\,,
where the vector field $X:\E\to\TE$ is the projection of $Y$
and\footnote{
Here $\n[o]$ is a shorthand for $\n[\Ch[o]\tn C]$\,,
the projection onto $\VU$ determined by the connection $\Ch[o]\tn C$\,.
} 
$\chY[o]\equiv\n[o]Y:\U\to\VU$
can be viewed as an anti-Hermitian endomorphism $\E\to\Lie$\,,
with coordinate expression
$$\chY[o]=\bigl(Y\Ii\sA\sB-\iO\,X^\l\,(
\Ch_\l\,\d\Ii\sA\sB-\oh\,C_\l^i\,\s\iIi i\sA\sB)\bigr)\,\zzB\tn\zeA~.$$
Conversely,
let $X:\E\to\TE$ be a vector field,
$\chY:\E\to\Lie$
an anti-Hermitian endomorphism and $o:\E\to\JE$ an observer.
Then one has the Hermitian linear projectable vector field
$$Y[X,o]\equiv X\pint(\Ch[o]\tn C)+\chY:\U\to\TO\U~,$$
with coordinate expression
$$Y[X,o]=
X^\l\,\de\xx_\l+\bigl(\iO\,X^\l\,(
\Ch_\l\,\d\Ii\sA\sB-\oh\,C_\l^i\,\s\iIi i\sA\sB)
+\chY\Ii\sA\sB\bigr)\,\zzB\,\dezA~.$$
\end{theorem}\qed

Let now \hb{$f{+}\phi:\JE\cart{\E}(\LL^*\tn\VE)\to\RR$}
be a special phase function for a classical spin particle
(\S\ref{ss:Special functions for a spin particle}).
Then $f:\JE\to\RR$ is a special phase function for a scalar particle,
and a geometric construction yields the vector field
\hb{$X[f]:\E\to\TE$}
(\S\ref{ss:Lie bracket of extended special functions})
and the real function $f[o]:\E\to\RR$
(the latter depending from the choice of an observer $o$).
Furthermore the linear function $\phi:\LL^*\tn\VE\to\RR$\,,
seen as a section $\E\to\LL\tn\VS\E$\,,
can be associated with the traceless anti-Hermitian endomorphism
$\Pau(\phi^\#):\E\to\Lie_0$\,,
where $\phi^\#\equiv g^\#(\phi)$\,.
Now $f[o]$ and $\Pau(\phi^\#)$ can be merged into a single section
$$\chY[f{+}\phi,o]\equiv\iO\,f[o]\,\id+\Pau(\phi^\#):\E\to
(\iO\,\RR\,\id)\dir{\E}\Lie_0=\Lie~.$$
Summarizing:
\begin{theorem}\label{functioncharacterization}
Let $f{+}\phi:\JE\cart{\E}(\LL^*\tn\VE)\to\RR$ be a special function
and $o:\E\to\JE$ an observer. Then $f{+}\phi$ is characterized
by a pair constituted by the vector field $X[f]$ and the
anti-Hermitian endomorphism $\chY[f{+}\phi,o]$\,,
whose coordinate expressions are
\begin{align*}
&X[f]=f^0\,\de\xx_0-f^j\,\de\xx_j~,
\\[10pt]
&\chY[f{+}\phi,o]=f[o]\,\id+\phi^i\,\xi_i=\\[6pt]
&\phantom{\chY[f{+}\phi,o]}
=\iO\,\bigl( f[o]\,\d\Ii\sA\sB
-\oh\,\phi^i\,\s\iIi i\sA\sB \bigr)\,\zzB\tn\zeA~,\quad
\phi^i\equiv g^{ij}\,\phi_j \equiv(\phi^\#)^i~,
\end{align*}
\end{theorem}\qed

Finally we join the results of the above two theorems.
The main result here is that by combining two observer-dependent isomorphisms
one gets an \emph{observer-independent} isomorphism
between special functions and Hermitian vector fields.
\begin{theorem}
Let $f{+}\phi:\JE\cart{\E}(\LL^*\tn\VE)\to\RR$ be a special function
and $o:\E\to\JE$ an observer.
Let \hb{$Y[f{+}\phi]:\U\to\TU$}
be the Hermitian linear projectable vector field obtained
from the pair \hb{$\bigl(X[f],\chY[f{+}\phi,o]\bigr)$}
via the correspondence stated in theorem~\ref{fieldcharacterization}.
Then $Y[f{+}\phi]$ turns out to be independent of $o$\,.
Conversely, let $Y:\U\to\TU$ be a Hermitian linear projectable vector field;
then the special function corresponding (via $o$)
to the pair $(X,\chY[o])$\,,
by virtue of theorem~\ref{functioncharacterization},
turns out to be independent of $o$\,.
Furthermore the above said operations are mutually inverse,
namely they determine an observer-independent one-to-one correspondence
\hb{$f{+}\phi\leftrightarrow Y[f{+}\phi]$}
between special functions and Hermitian vector fields.
This correspondence turns out to be an isomorphism of Lie algebras,
the product of vector fields being given by standard Lie bracket
and that of special functions by the bracket introduced
in~\S\ref{ss:Lie bracket of extended special functions}.
\end{theorem}
\proof
In coordinates adapted to $o$ we find
\begin{align*}
Y[f{+}\phi]&=f^0\,\de\xx_0-f^j\,\de\xx_j+
\\[6pt]&\phantom={}+
\Bigl( \iO\,(f^0A_0-f^jA_j+\br f)\,\d\Ii\sA\sB-
\ih\,(f^0\,C_0^i-f^j\,C_j^i+\phi^i)\,\s\iIi i\sA\sB
\Bigr)\,\zzB\,\dezA~,
\end{align*}
that is
$Y[f{+}\phi]=X^\l\,\de\xx_\l+Y^\l\,\xi\iIi\l\sA\sB\,\zzB\,\dezA$
with
$$X^0=f^0~,~~X^i=-f^i~,~~Y^0=f^0A_0-f^jA_j+\br f~,~~
Y^i=X^\l\,C_\l^i+\phi^i\equiv X^l\,C_\l^i+g^{ij}\,\phi_j~.$$
Now, while $A_\l$ and $\br f$ depend on the observer,
the combination \hbox{$f^0A_0{-}f^jA_j{+}\br f$} can be shown to be
independent of the observer~\cite{JM06}.
Thus $Y[f{+}\phi]$ turns out to be observer-independent.
By a direct calculation,
the correspondence \hb{$f{+}\phi\leftrightarrow Y[f{+}\phi]$}
can be checked to be an isomorphism of Lie algebras
\qed

\remark~If \hb{$\chY[o]:\E\to\Lie$} is the endomorphism
determined by the Hermitian linear projectable vector field
$Y:\U\to\TU$ via the connection $\Ch[o]\tn C$\,,
then one gets the scalar case special function $f:\JE\to\RR$
characterized by \hb{$f^0=X^0$}\,, \hb{$f^i=-X^i$} and
\hb{$\br f=-\ih\,\Tr\chY[o]$}\,,
while the traceless part of $\chY[o]$ yields,
via the Pauli map and the spacelike metric,
the section \hb{$\phi:\E\to\LL\tn\VS\E$}\,.\rembox

%--------------
\subsection{Quantum operators}\label{ss:Quantum operators}

We conclude with a brief sketch of the main ideas
concerning the relation between special functions,
Hermitian vector fields and quantum operators
(see~\cite{JM06} for some recent references on the scalar case;
for the spin case this topic is essentially treated in~\cite{CJM95},
though it would require some updating in order to cope with the most
recent results).

We first introduce a modification
of the setting of the previous sections:
we assume the Hermitian product $h$ of the fibers of $\U$
to be scaled, namely a section
\hb{$h:\E\to\LL^{-3}\tn\Ua\ten{\E}\Ul$}\,.
Since the volume form \hb{$\eta:\E\to\LL^3\tn\weu3\VS\E$}
associated with the spacelike metric is $\LL^3$-scaled,
we get an \emph{unscaled} section
$$h\tn\eta:\E\to\Ua\ten{\E}\Ul\ten{\E}\weu3\VS\E~.$$

For all $t\in\T$ let $\DCo(\E_t,\U\!_t)$ denote the vector space
of all $\CO^\infty$ sections $\E_t\to\U\!_t$ with compact support.
This space has a standard topology~\cite{Sc};
using $h\tn\eta$\,,
its topological dual  $\DC(\E_t,\U\!_t)$ can be identified with the
vector space of all \emph{generalized sections}
(in the distributional sense) of the same bundle.
A locally integrable ordinary section $\theta:\E_t\to\U\!_t$, in particular,
can be identified with the element in $\DC(\E_t,\U\!_t)$ acting as
$$\bang{\theta,\phi}:=
\int_{\E_t}h(\theta,\phi)\,\eta~,\quad \phi\in\DCo(\E_t,\U\!_t)~.$$

Consider, moreover, the vector space
$$\LC^2(\E_t\,,\,\U\!_t)
:=\{\psi_t:\E_t\to\U\!_t:\bang{\psi_t\,,\,\psi_t}<\infty\}~.$$
This yields, by the usual procedure,
a Hilbert space $\HC(\E_t,\U\!_t)$\,,
constituted by equivalence classes of elements in $\LC^2(\E_t\,,\,\U\!_t)$
differing on zero-measure sets.
We have natural inclusions
$$\DCo(\E_t,\U\!_t)\subset\HC(\E_t,\U\!_t)\subset\DC(\E_t,\U\!_t)~,$$
namely a so-called \emph{rigged Hilbert space}~\cite{BLT}.
Furthermore, the fibered sets
$$\DCo(\E,\U):=\bigsqcup_{t\in\T}\DCo(\E_t\,,\,\U\!_t)~,~~
\HC(\E,\U):=\bigsqcup_{t\in\T}\HC(\E_t\,,\,\U\!_t)~,~~
\DC(\E,\U):=\bigsqcup_{t\in\T}\DC(\E_t\,,\,\U\!_t)~,$$
have a natural vector bundle structure over $\T$, given by
Fr\"olicher's notion of smoothness~\cite{Fr,MK,C04}.
So we get a \emph{rigged Hilbert bundle},
namely the sequence \hb{$\DCo\subset\HC\subset\DC$} of monomorphisms over $\T$.

Let \hb{$X:\E\to\TE$} be a vector field.
Then it is not difficult to see
that the Lie derivative $X.\a$ is well defined
for each $\a:E\to\weu3\VS\E$ iff\footnote{
If $\td\a:\E\to\weu3\TS\E$ is any extension of $\a$
then the vertical (spacelike) restriction o $X.\td\a$
is a section $E\to\weu3\VS\E$
if $X$ is projectable in the above sense.} 
$X$ is projectable through $\E\to\T$
(namely \hb{$X^0:\T\to\RR$}, see the concluding remark
of~\S\ref{ss:Special phase functions for a scalar particle}).
In that case
(which encompasses the cases of physical interest)
we may introduce
a slightly modified notion of a Hermitian linear projectable vector field
as preserving $h\tn\eta$ rather than $h$\,,
namely through the condition
\hb{$X.(h(\phi,\psi)\,\eta)=h(Y.\phi,\psi)\,\eta+h(\phi,Y.\psi)\,\eta$}\,.
In coordinates
$$h_{\cA\sB}\,\bar Y\Ii\cA\cB+h_{\cB\sA}\,Y\Ii\sA\sB+
\frac1{\sqrt{|g|}}\,\bigl(X^0\,\de_0\sqrt{|g|}+\de_i(X^i\,\sqrt{|g|})\bigr)
=0~,$$
where the new term
\hb{$\bigl(X^0\,\de_0\sqrt{|g|}+\de_i(X^i\,\sqrt{|g|})\bigr)/\sqrt{|g|}
=\bang{X.\eta\,,\eta^{-1}} \equiv\diveta X$}
can be interpreted as the divergence of $X$ relatively to $\eta$\,.
Again one finds that local Hermitian projectable vector fields
constitute a sheaf of Lie algebras.
The results of the previous sections hold essentially unchanged,
with $\chY[o]$ modified by the term $\oh\,(\diveta X)\,\id$\,.

We now observe that an ordinary section \hb{$\psi:\E\to\U$}
can be identified with a section \hb{$t\mapsto\psi_t$}
of some functional bundle over $\T$.
Conversely, a section (say) \hb{$\T\to\DCo(\E,\U)$} can be seen as a section
\hb{$\E\to\U$}\,.
Moreover we recall (\S\ref{ss:Hermitian vector fields})
that a linear projectable vector field acts on sections \hb{$\psi:\E\to\U$}
by a natural generalization of the standard Lie derivative.
Hence each special function $f{+}\phi$ acts on $\psi$ as
\hb{$Y[f{+}\phi].\psi$}\,.
Then it's quite natural to ask when does this operation yield
a fibered action on $\DCo\equiv\DCo(\E,\U)$ over $\T$.
The answer is that it must not involve the derivative of $\psi$
relatively to the ``timelike'' coordinate $\xx^0$.
In those cases when this condition is not fulfilled, however,
one sees that there is a unique linear combination
of the above said operation and the \emph{Pauli operator}\footnote{
This is a straightforward generalization of the \emph{Schr\"odinger operator}
of the scalar case~\cite{JanMod02c},
which allows replacing the time derivative
with a Laplace-type operator on the fibers of \hb{$\E\to\T$}\,.
See~\S\ref{ss:Main examples} for its explicit
coordinate expression.
Essentially, the Pauli operator arises as the Euler-Lagrange operator
of the quantum Lagrangian
for the spin particle~\cite{CJM95}.} 
\hb{$\mathfrak{P}:\JO_2\U\to\TT^*\tn\U$}
by which one can ``eliminate'' the time derivative;
we then get the ``pre-quantum'' fibered operator
$$\widehat{f{+}\phi}:\DCo\to\DCo:\psi\mapsto
\iO\,\bigl(Y[f{+}\phi].\psi-u_0\,f^0\,\mathfrak{P}\psi \bigr)~,$$
which turns out to be symmetric
(hence self-adjoint, given suitable domain conditions).
Note that
\emph{this is the essential reason for considering Hermitian vector fields:
eventually obtaining self-adjoint operators}.

Pre-quantum operators are naturally extended
to linear fibered operators \hb{$\DC\to\DC$}\,,
while extension to fibered \emph{quantum operators}
\hb{$\HC\to\HC$} on the Hilbert bundle is not guaranteed.

The bracket of two pre-quantum operators is defined by
$$[\Ycal,\Ycal']:=-\iO\,(\Ycal\comp\Ycal'-\Ycal'\comp\Ycal)~.$$
Then we get a Lie-algebra isomorphism
between special functions and pre-quantum operators.

\remark~In the particular case of special functions
which do not contain the energy
(namely special functions such that \hb{$f^0=0$})
we simply have \hb{$(\widehat{f{+}\phi})\psi=\iO\,Y[f{+}\phi].\psi$}\,.~\remend

%--------------
\subsection{Main examples}\label{ss:Main examples}

The Pauli operator \hb{$\mathfrak{P}:\JO_2\U\to\TT^*\tn\U$}
acts on a local section \hbox{$\psi=\psi^\sA\,\zeA:\E\to\U$} as
$$\mathfrak{P}[\psi]^\sA=
\mathfrak{S}[\psi^\sA]-u^0\,C\iIi0\sA\sB\,\psi^\sB~;$$
here $\mathfrak{S}$ is the \emph{Schr\"odinger operator}\,,
acting on scalar sections $\breve\psi:\E\to\Q$ as
$$\mathfrak{S}[\breve\psi]=
u^0\,\bigl(\de_0-\iO\,A_0+\frac{\de_0\sqrt{|g|}}{2\,\sqrt{|g|}}
-\frac\iO2\,\Delta_0[o]\bigr)\breve\psi~,$$
where the \emph{observed Laplacian} $\Delta_0[o]$\,,
relatively to the observer $o:\E\to\JE$
(\S\ref{ss:The phase space of a classical particle}),
has the coordinate expression
$$\Delta_0[o]=u_0\,\frac\h m\,g^{ij}\,\bigl(
(\de_i-\iO\,A_i)\,(\de_j-\iO\,A_j)+K\iIi ihj\,(\de_h-\iO\,A_h)\bigr)~.$$

The main instances of special functions are:
the coordinates $\xx^\l$\,;
the classical momentum components
\hbox{$\Pcal_i=\frac{m\,u^0}{\h}\,g_{ij}\,\xx_0^j+A_i$}\,;
the classical Hamiltonian
\hbox{$\Hcal_0'=\Hcal_0-u_0\,\m\,B^\fl$}\,,
where
\hbox{$\Hcal_0=\frac{m\,u^0}{2\,\h}\,g_{ij}\,\xx_0^i\,\xx_0^j-A_0$}
is the scalar case Hamiltonian and
\hbox{$-u_0\,\m\,B^\fl:\E\to\LL\tn\VS\E$}
(a real-valued function on the spin phase space)
corresponds to $u_0/\h$ times the energy of interaction between
the particle's spin and the magnetic field
(see \S\ref{ss:Special functions for a spin particle} and
footnote~\ref{foot:magneticfield} on page~\pageref{foot:magneticfield});
and, finally, the \emph{spin in the $n$ direction},
$n^\fl\equiv g^\fl(n)=n_i\,\check{\dx}{}^i$\,,
where $n:\E\to\LL\tn\VS\E$ is a \emph{unit} covector field.
Then we find
\begin{align*}
\widehat{\xx^\l}\psi&=\xx^\l\,\psi
\\[6pt]
\widehat{\Pcal_i}\psi&=-\iO\,\bigl(\na_i+\frac{\de_i\sqrt{|g|}}{2\,\sqrt{|g|}}
\bigr)\,\psi=
-\iO\,\bigl(\de_i\psi^\sA-C\iIi i\sA\sB\,\psi^\sB
+\frac{\de_i\sqrt{|g|}}{2\,\sqrt{|g|}}\,\psi^\sA\bigr)\,\ze_\sA
\\[6pt]
\widehat{\Hcal_0'}\psi&=
\bigl(-\oh\,\Delta_0[o]-A_0+\iO\,u_0\,\m\,\Pau[B]\bigr)\psi=
\bigl(
-\oh\,\Delta_0[o]\psi^\sA-A_0\,\psi^\sA
+\oh\,u_0\,\m\,B^i\,\s\iIi i\sA\sB\,\psi^\sB\bigr)\,\zeA~,
\\[6pt]
\widehat{n^\fl}\psi&=\iO\,\Pau[n]\psi
=\oh\,n^i\,\s\iIi i\sA\sB\,\psi^\sB\,\zeA~.
\end{align*}

%==============================================================================%

%==============================================================================%

\end{document}